\documentclass[11pt, oneside]{article}   	
\usepackage{geometry}                		
\usepackage{graphicx}				
\usepackage{amssymb}

\usepackage{amsmath}

\usepackage{endnotes}
\let\footnote=\endnote

\title{Ammonia Inversion Energy Levels using Operator Algebra}
\author{S M Blinder\footnote{email: sblinder@wolfram.com and sblinder@umich.edu}   \\  \\
Wolfram Research Inc., Champaign, IL  61820-7237 USA \\ and \\ University of Michigan, Ann Arbor, MI 48109-1055 USA}
\date{}							

\begin{document}
\maketitle

\centerline{\Large\bf Abstract}

\vspace{.25cm}

\noindent The inversion potential for the ammonia molecule is approximated by $V(x)=k(x^2-r^2)^2/8r^2$. The Hamiltonian thereby contains only even powers of $p$ and $x$ and a representation in terms of ladder operators $a$ and $a^\dagger$ is suggested. The frequency variable $\omega$  occurring in the operators is introduced as a free parameter, with its value to be determined such as to optimize agreement with experimental results. Using  known structural parameters of the ammonia molecule, the eigenvalues for a $10\times 10$ truncation of the Hamiltonian matrix are computed. The splitting between the two lowest eigenvalues corresponds to the ammonia maser frequency,  24.87 GHz, this value being reproduced by the appropriate choice of $\omega$. \\ \\

\begin{figure}
\begin{center}
\includegraphics[height=5cm]{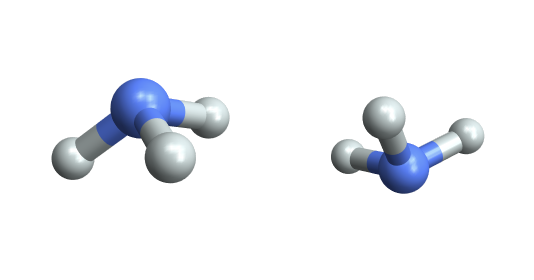}
\caption{Ammonia molecule in its two metastable pyramidal states.}
\label{am}
\end{center}
\end{figure}

\begin{figure}
\begin{center}
\includegraphics[height=5cm]{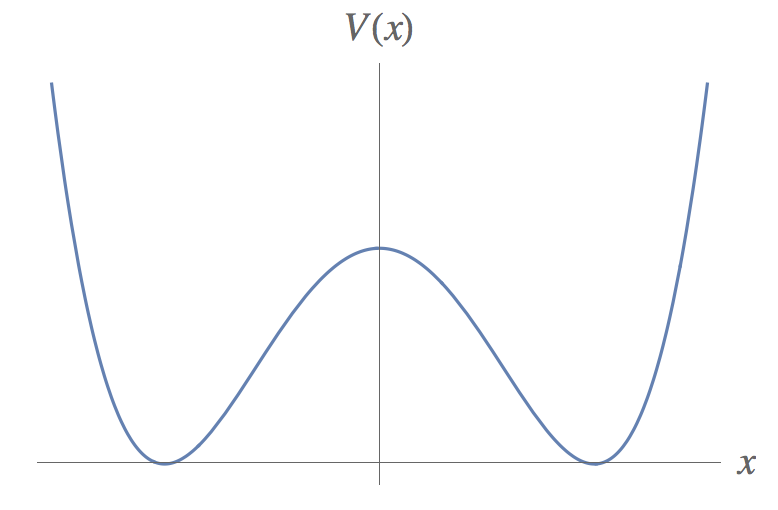}
\caption{Ammonia inversion potential.}
\label{Vx}
\end{center}
\end{figure}

The inversion of the ammonia molecule NH$_3$, shown in Fig. \ref{am}, can be described by a simplified Schr\"odinger equation
\begin{equation}\label{H}
\frac{p^2}{2\mu}\psi(x)+V(x)\psi(x)=\varepsilon\psi(x).
\end{equation}
Here only the linear motion of the nitrogen atom is considered, with neglect of the other vibrational modes of the molecule. Here $\mu$ is the reduced mass of the nitrogen molecule, given by
\begin{equation}\label{mu}
\mu=\frac{3m_N m_H}{m_N+3m_H}.
\end{equation}
A number of analytic representations of the inversion potential were discussed by Swalen and Ibers.\footnote{J. D. Swalen and J. A. Ibers, ``Potential Function for the Inversion of Ammonia,'' {\it J. Chem. Phys.} {\bf 36}(7) (1962), pp. 1914-1918.}
We consider a potential of the form (Fig. \ref{Vx}):
\begin{equation}\label{V}
V(x)=\frac{k(x^2-r^2)^2}{8r^2} = \frac{k r^2}{8}-\frac{k x^2}{4}+\frac{k x^4}{8r^2},  
\end{equation}
with minima at $x=\pm r$ and a barrier of height $k r^2/8$ at $x=0$. This general form was utilized by Damburg and Propin.\footnote{R. J. Damburg and R. Kh. Propin, ``Model Potential for Inversion in Ammonia,'' {\it Chemical Physics Letters} {\bf 14}(1) (1972), pp. 82-84; the form of the potential was first introduced by Certain, Hirschfelder, Kolos, and Wolniewicz, in a study of exchange interactions in the H$_2^+$ molecule.}

Since the Hamiltonian contains only even powers of $p$ and $x$, a representation based on the ladder operators $a$ and $a^\dagger$ suggests itself, a generalization of the canonical operator formulation for the harmonic oscillator. Accordingly, we define
\begin{equation}
a = \sqrt{\frac{\mu\omega}{2}} x + i\sqrt{\frac{1}{2 \mu \omega}} p, \qquad
a^\dagger = 
 \sqrt{\frac{\mu\omega}{2}} x - i \sqrt{\frac{1}{2 \mu \omega}} p.
\end{equation}
The parameter $\omega$ is introduced, with its value to be determined such as to optimize agreement with experimental results.
The actions of the ladder operators on a basis ket are given by
\begin{equation} 
a|n\rangle = \sqrt{n}|n-1\rangle,  \qquad a^\dagger |n\rangle = \sqrt{n+1}|n+1\rangle.
\end{equation}

The Hamiltonian in Eq. (\ref{H}), can be expanded  to give
\begin{equation}
H= \frac{p^2}{2\mu}+\frac{k r^2}{8}-\frac{k}{4} x^2+\frac{k}{8r^2}x^4.
\end{equation}
In terms of the ladder operators, we have
\begin{equation}
x = \sqrt{\frac{1}{2 \mu \omega}} (a + a^\dagger), \qquad 
p = -i \sqrt{\frac{\mu \omega}{2}} (a - a^\dagger),
\end{equation}
so that
\begin{equation}
x |n\rangle =  \sqrt{\frac{1}{2 \mu \omega}} \Big(\sqrt{n}|n-1\rangle + \sqrt{n+1}|n+1\rangle\Big)
\end{equation}
and
\begin{equation}
p |n\rangle= -i \sqrt{\frac{\mu \omega}{2}} \Big(\sqrt{n}|n-1\rangle - \sqrt{n+1}|n+1\rangle\Big).
\end{equation}
By successive application of these operators, it follows that
\begin{equation}
x^2 |n\rangle =  \frac{1}{2 \mu \omega} \Big(\sqrt{n(n-1)} |n-2\rangle +(2n+1) |n\rangle +\sqrt{(n+1)(n+2)} |n+2\rangle \Big)
\end{equation}
and
\begin{equation}
p^2 |n\rangle =  -\frac{\mu \omega}{2} \Big(\sqrt{n(n-1)} |n-2\rangle -(2n+1) |n\rangle +\sqrt{(n+1)(n+2)} |n+2\rangle \Big).
\end{equation}
Note, incidentally, that
\begin{equation}
\Big(\frac{p^2}{2\mu}+\frac{1}{2}\mu\omega^2 x^2 \Big) |n\rangle =\left(n+\frac{1}{2}\right)\omega |n\rangle,
\end{equation}
which agrees with the result for an harmonic oscillator. Finally, we need
\begin{eqnarray}
x^4 |n\rangle =  \frac{1}{4 \mu^2 \omega^2} \Big(\sqrt{n(n-1)(n-2)(n-3)} |n-4\rangle+
2\sqrt{n(n-1)}(2n-1) |n-2\rangle+ \hspace{1cm} \nonumber \\
(6n^2+6n+3)  |n\rangle+ 
2\sqrt{(n+1)(n+2)} (2n+3)  |n+2\rangle+ \hspace{1cm} \nonumber \\
\sqrt{(n+1)(n+2)(n+3)(n+4)} |n+4\rangle \Big). \hspace{1cm}
\end{eqnarray}

The nonzero matrix elements  of the Hamiltonian are given by
\begin{equation}
H_{n,n}= \frac {1} {32} \Big (\frac {(6 n^2 + 6 n + 
        3) k} {r^2 \mu^2 \omega^2} + 
   4 k r^2 - \frac {(8 n + 4) k} {\mu\omega} + (16 n + 8)\omega \Big),
\end{equation}
\begin{equation}
H_{n+2,n}= \frac {\sqrt {(n + 1) (n + 2)}\Big (-4 r^2 \mu^2 \omega^3 + (2 n + 
        3 - 2 r^2\mu\omega)k \Big)} {16 r^2 \mu^2\omega^3},
\end{equation}
\begin{equation}
H_{n+4,n}= \frac{\sqrt{(n+1)(n+2)(n+3)(n+4)}k}{32r^2 \mu^2\omega^2}.
\end{equation}
The matrix is symmetrical, so that $H_{m,n}=H_{n,m}$.

For explicit computation, we need numerical values for the parameters $\mu$, $r$ and $\omega$. We  use Hartree atomic units with $\hbar = m_e =e =1$. The mass of a hydrogen atom is given by $m_H =1837.153$, and a nitrogen atom by $m_N=25530.80$. Thus the reduced mass of the nitrogen atom, as it participates in  inversion (Eq. \ref{mu}),  is equal to $\mu= 4532.92$. The experimentally-determined equilibrium displacement of the nitrogen atom from the plane of the three hydrogen atoms gives $r$= 0.3816 \AA= 0.7211 bohr.\footnote{National Institute of Standards and Technology, tabulation at https://cccbdb.nist.gov/expdata.asp.}
The accepted value for the inversion barrier for ammonia is $V_0$=24.2 kJ/mol = 0.009243 hartrees.\footnote{A. Rauk and L. C. Allen,``Electronic Structure and Inversion Barrier of Ammonia,'' {\it J. Chem. Phys.} {\bf 52}(8) (1970) pp. 4133-4144.}
We assign the value $\omega = 0.00206226$.

The 10$\times$10 truncated matrix of $H$ is shown below:
\begin{figure}[h]
\begin{center}
\includegraphics[height=3cm]{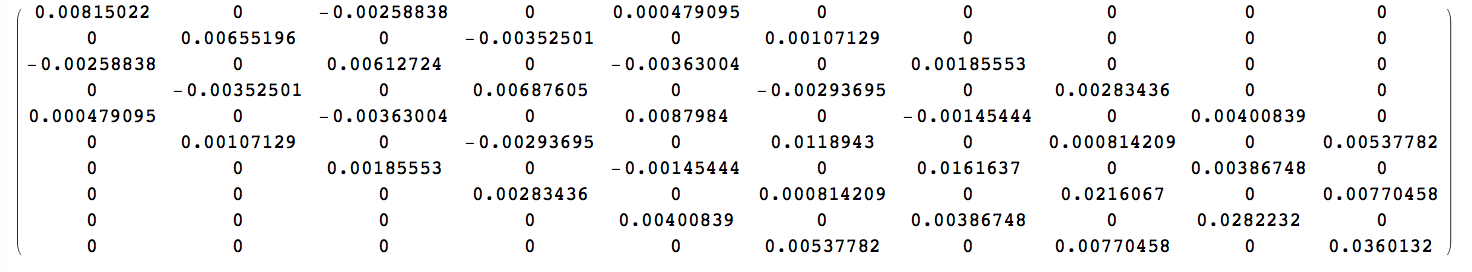}
\caption{Matrix $H_{mn}$.}
\label{mx}
\end{center}
\end{figure}

\begin{figure}[h]
\begin{center}
\includegraphics[height=5cm]{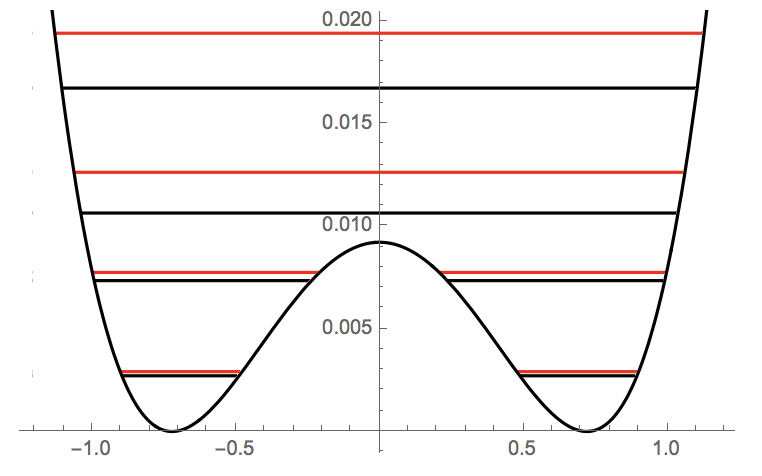}
\caption{Computed eigenvalues plotted with inversion potential. Values in atomic units.}
\label{evals}
\end{center}
\end{figure}

\noindent The eigenvalues are calculated using the {\bf Eigenvalue} program in 
{\it Mathematica}\textsuperscript{TM}.
The lowest eight eigenvalues are given by
\begin{eqnarray}
\varepsilon_0=0.00272086,\ \varepsilon_1=0.00272464,\ \varepsilon_2= 0.00736816,\ \varepsilon_3=0.00776894, \nonumber \\   \varepsilon_4=0.0106695,  \varepsilon_5= 0.0126591,\ \varepsilon_6=0.0167717,\ \varepsilon_7= 0.0194514.
\end{eqnarray}
These occur in closely-spaced pairs, representing nearly-degenerate symmetric and antisymmetric states.
The eigenvalues are plotted in Fig. \ref{evals},  with the symmetric and antisymmetric states are shown in black and red, respectively. The computed splitting in the ground torsional state is given by $\varepsilon_1-\varepsilon_0 = 3.7941 \times 10^{-6}$ hartrees = 24.87 GHz, the frequency of the ammonia maser. The value of $\omega$ has been adjusted to produce this agreement.

\vspace{1cm}

\leftline{\Large\bf References}

\theendnotes

\end{document}